\documentclass[conference]{IEEEtran}
\IEEEoverridecommandlockouts
\usepackage{cite}
\usepackage{amsmath,amssymb,amsfonts}
\usepackage{algorithmic}
\usepackage{graphicx}
\usepackage{textcomp}
\usepackage{xcolor}
\usepackage{bbm}

\usepackage[textsize=tiny]{todonotes}

\def\BibTeX{{\rm B\kern-.05em{\sc i\kern-.025em b}\kern-.08em
    T\kern-.1667em\lower.7ex\hbox{E}\kern-.125emX}}

\DeclareMathOperator*{\argmax}{\mathrm{argmax}}
\begin{document}

\title{Experimental Investigation of Deep Learning for \\ Digital Signal Processing in Short Reach \\ Optical Fiber Communications}

\newcommand\blfootnote[1]{%
  \begingroup
  \renewcommand\thefootnote{}\footnote{#1}%
  \addtocounter{footnote}{-1}%
  \endgroup
}

\author{
  \IEEEauthorblockN{
    Boris Karanov\IEEEauthorrefmark{1}\IEEEauthorrefmark{3},
    Mathieu Chagnon\IEEEauthorrefmark{3},
    Vahid Aref\IEEEauthorrefmark{3},
    Filipe Ferreira\IEEEauthorrefmark{1},
    Domani\c{c} Lavery\IEEEauthorrefmark{1},
    Polina Bayvel\IEEEauthorrefmark{1}, \\
    and Laurent Schmalen\IEEEauthorrefmark{2}
  }
  \vspace*{1ex}
  \IEEEauthorblockA{
    \IEEEauthorrefmark{1}Dept. Electronic \& Electrical Engineering, University College London, WC1E 7JE London, U.K. 
  }
  \IEEEauthorblockA{
    \IEEEauthorrefmark{2}Communications Engineering Lab, Karlsruhe Institute of Technology (KIT), 76131 Karlsruhe, Germany
  }
  \IEEEauthorblockA{
    \IEEEauthorrefmark{3}Nokia Bell Labs, 70435 Stuttgart, Germany
  }
}

\IEEEspecialpapernotice{(Invited Paper)}

\maketitle

\begin{abstract}
We investigate methods for experimental performance enhancement of auto-encoders based on a recurrent neural network~(RNN) for communication over dispersive nonlinear channels. In particular, our focus is on the recently proposed sliding window bidirectional RNN (SBRNN) optical fiber auto-encoder. We show that adjusting the processing window in the sequence estimation algorithm at the receiver improves the reach of simple systems trained on a channel model and applied ``as is'' to the transmission link. Moreover, the collected experimental data was used to optimize the receiver neural network parameters, allowing to transmit 42\,Gb/s with bit-error rate (BER) below the 6.7\% hard-decision forward error correction threshold at distances up to 70\,km as well as 84\,Gb/s at 20\,km. The investigation of digital signal processing (DSP) optimized on experimental data is extended to pulse amplitude modulation with receivers performing sliding window sequence estimation using a feed-forward or a recurrent neural network as well as classical nonlinear Volterra equalization. Our results show that, for fixed algorithm memory, the DSP based on deep learning achieves an improved BER performance, allowing to increase the reach of the system.
\end{abstract}


\section{Introduction}\label{sec:Intro}
\blfootnote{This work was funded by the European Union’s Horizon 2020 research and
innovation programme under the Marie Skłodowska-Curie project COIN (No. 676448) and the UK EPSRC programme grant TRANSNET (EP/R035342/1).} The application of machine learning in the design, networking and monitoring of communication systems has attracted great research interest in the recent years~\cite{Simeone,Rafique,Jiang,Khan}. From a physical layer perspective, deep learning techniques~\cite{Goodfellow}, enabled by artificial neural networks~(ANN), are considered a viable alternative to classical digital signal processing (DSP)~\cite{Simeone,Khan}. The combination of ANNs and deep learning can be used to optimize DSP functions at the transmitter or receiver. For example, deep learning was applied to the digital back-propagation block for non-linearity compensation in long-haul coherent optical fiber systems, reducing the computational demands~\cite{Hager}. Furthermore, an ANN was used as a receiver equalizer for low-cost fiber-optic communication links as found in passive access networks (PONs)~\cite{Houtsma}. In molecular communications, a novel detection scheme based on neural networks in combination with an efficient sliding window sequence estimation algorithm was proposed in~\cite{Farsad}. The aforementioned applications of deep learning are examples where the techniques are within a specific DSP module. The signal processing for such communication systems is engineered and optimized module-by-module, each part being responsible for an individual task such as coding, modulation or equalization.

\begin{figure*}[t!]
\centering
\includegraphics[width=\textwidth, keepaspectratio=true]{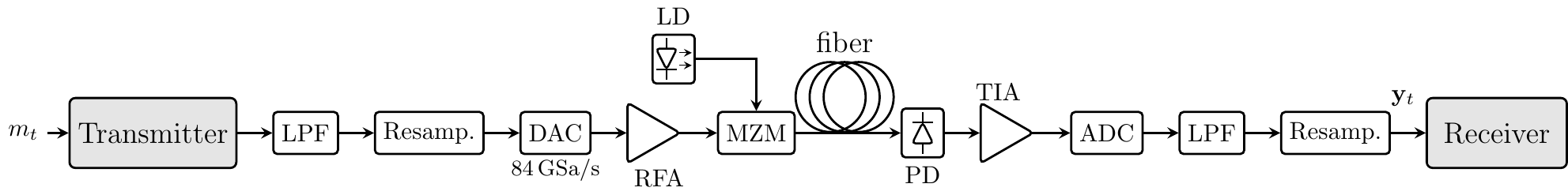}\vspace*{-1ex}
\caption{\label{fig:experiment_schematics} Schematic of the experimental optical IM/DD transmission test-bed used for investigation of the digital signal processing schemes.}\vspace*{-2ex}
\end{figure*}

A different approach to deep learning-based DSP, which utilizes the full potential of the function approximation capabilities of artificial neural networks, is to interpret the complete communication chain from the transmitter input to the receiver output as a single deep ANN~\cite{O'Shea,Doerner} and optimize its parameters in an end-to-end process. Such fully learnable communication transceivers (also known as \emph{auto-encoders}) partly avoid the modular design of conventional systems and have the potential to achieve an end-to-end optimized performance for a specific metric. Auto-encoders are especially suitable for communication over channels where the optimum transceiver is unknown or its implementation is computationally prohibitive. Dispersive nonlinear channels are an example, which is often encountered in optical fiber communications. This prompted the introduction of the auto-encoder concept for such systems~\cite{Karanov_1,Li,Jones}. In particular, end-to-end deep learning is seen as a viable DSP solution to enhance the system performance at reduced complexity in low-cost intensity modulation/direct detection (IM/DD) optical fiber links where the combination of chromatic dispersion, introducing inter-symbol interference (ISI), and nonlinear signal detection by a photo-diode imposes severe limitations~\cite{Agrawal,Chagnon_2}. The first experimental demonstration of an optical fiber auto-encoder showed that a simple feed-forward neural network (FFNN) design for IM/DD can outperform pulse amplitude modulation (PAM) with conventional, simple linear equalizers~\cite{Karanov_1}. Addressing the limitations of the FFNN system in handling the channel memory, a transceiver based on a bidirectional recurrent ANN and sliding window sequence estimation (SBRNN) was proposed in~\cite{Karanov_2} and experimentally verified in~\cite{Karanov_4}, achieving a substantial performance improvement. Moreover, the SBRNN auto-encoder outperformed PAM transmission with state-of-the-art nonlinear receivers~\cite{Karanov_2} or classical maximum likelihood sequence detection (MLSD)~\cite{Karanov_3} without the associated computational complexity of such schemes.

In this paper, we extend the experimental study of the SBRNN auto-encoder by investigating methods to further enhance the performance it achieves when parameters are trained on a channel model and applied to the transmission link. A simple solution, which does not require the implementation of an additional training procedure, is to adjust the sliding window size in the estimation algorithm and thus capture more of the channel memory. Moreover, we show that the BER can be improved by initiating an optimization for the receiver parameters using the collection of experimental data. In this way, the performance penalty due to any mismatch between the simplified channel model and the actual transmission link can be reduced. We extend the investigation of DSP schemes using experimental data and deep learning by considering PAM systems with sliding window FFNN or SBRNN receivers and comparing them to a classical nonlinear Volterra equalizer. While the SBRNN auto-encoder had a superior performance, the SFFNN was shown as a viable receiver alternative for PAM systems with an improved BER over Volterra equalizers.

\section{Experimental Setup and Data Collection}\label{sec:Exp_setup}
Due to its simplicity and cost-effectiveness, optical fiber transmission based on IM/DD is considered a front-runner technology for \emph{fiber to the x} (FTTX) systems. The IM/DD links are also preferred for other short reach applications such as data center, metro and access networks~\cite{Chagnon_2}. In this section we describe the experimental IM/DD test-bed used to examine the performance of the DSP schemes and explain how the data-sets for parameter optimization and testing were formed.
\subsection{Optical Fiber Transmission Link}\label{sec:transmission_link}
Figure~\ref{fig:experiment_schematics} shows a schematic of the test-bed. We consider an optically un-amplified transmission link. The output of the transmitter is filtered in the digital domain by a Brick-wall low-pass filter (LPF) which has a bandwidth of 32\,GHz. Afterwards, the signal is appropriately re-sampled (module \emph{Resamp.}) and applied to the digital-to-analog converter (DAC) which operates at 84\,GSa/s. The obtained electrical waveform is pre-amplified by a driver amplifier and used to modulate the 1550\,nm laser diode (LD) via a Mach-Zehnder modulator (MZM). For the auto-encoder system, the MZM is meticulously biased to match simulation, while for the PAM transmission, the bias is set at the quadrature point. The optical power input to the fiber is set to 1\,dBm. The modulated optical signal is propagated over a fixed length of standard single mode fiber and the received waveform is direct-detected by a PIN photo-diode (PD) combined with a trans-impedance amplifier (TIA). The signal is real-time sampled by an analog-to-digital converter (ADC). After synchronization, proper scaling, offset and re-sampling, the digitized photo-current is stored for the subsequent signal processing at the receiver.

\subsection{Data Collection for the Auto-encoder System}\label{sec:autoenc_data}
For the auto-encoder experiment we transmit $Z=800$ different sequences of $T=5152$ random input messages $m_{i,j}\!\in\!\mathcal{M}$ (with an alphabet size of $M\!=\!|\mathcal{M}|\!=\!64$), with $i\in\{1,\ldots,Z\}$ and $j\in\{1,\ldots,T\}$. The data is generated using the Mersenne twister algorithm to avoid learning representations of a pseudo-random sequence during receiver optimization~\cite{Eriksson}. Each of the sequences is first encoded by the transmitter BRNN into series of blocks of $n$ samples before being applied to the experimental test-bed, starting with the LPF. Note that we used an up-sampling factor of~4 in both transmitter and receiver. The value of $n$ depends on the data rate. For 42\,Gb/s or 84\,Gb/s transmission, we select $n\!=\!48$ or $n\!=\!24$, respectively.  Since the BRNN transmitter encodes a message carrying $\log_2M\!=\!6$~bits, the series of encoded blocks were down-sampled in experiment after the LPF such that 6~bits were carried by~12 or~6 DAC samples for 42\,Gb/s or 84\,Gb/s transmission, respectively. A full load of the DAC requires the concatenation of 8 encoded sequences and this is fed to the link for a single transmission iteration. To collect a sufficient amount of experimental data, we performed 100 DAC loads. For receiver optimization and testing we used the sequences of blocks $\mathbf{y}_{i,j}$, $i\in\{1,\ldots,Z\}$ and $j\in\{1,\ldots,T\}$, at the output of the test-bed together with their labels $m_{i,j}$. Sequences $i\in\{1,\ldots,720\}$ formed the training sets 
\begin{equation*}
\mathbf{D}^{\textnormal{AE}}:= 
\begin{bmatrix}
\mathbf{y}_{1,1\ldots T} & \mathbf{y}_{2,1\ldots T} & \mathbf{y}_{3,1\ldots T} & \mathbf{y}_{4,1\ldots T} \\
\mathbf{y}_{5,1\ldots T} & \mathbf{y}_{6,1\ldots T} & \mathbf{y}_{7,1\ldots T} & \mathbf{y}_{8,1\ldots T} \\
\vdots & \vdots & \vdots & \vdots \\
\mathbf{y}_{717,1\ldots T} & \mathbf{y}_{718,1\ldots T} & \mathbf{y}_{719,1\ldots T} & \mathbf{y}_{720,1\ldots T}
\end{bmatrix}
\end{equation*}
\begin{equation*}
\mathbf{L}^{\textnormal{AE}}:= 
\begin{bmatrix}
m_{1,1\ldots T} & m_{2,1\ldots T} & m_{3,1\ldots T} & m_{4,1\ldots T} \\
m_{5,1\ldots T} & m_{6,1\ldots T} & m_{7,1\ldots T} & m_{8,1\ldots T} \\
\vdots & \vdots & \vdots & \vdots \\
m_{717,1\ldots T} & m_{718,1\ldots T} & m_{719,1\ldots T} & m_{720,1\ldots T}
\end{bmatrix},
\end{equation*}
where $\mathbf{D}^{\textnormal{AE}}\!\in\!{\mathbb{R}}^{180\times 4\cdot T\cdot n}$ consists of the data elements $\mathbf{y}_{i,j}$ and $\mathbf{L}^{\textnormal{AE}}\!\in\!{\mathbb{R}}^{180\times 4\cdot T}$ contains the label elements $m_{i,j}$. During the training procedure (see Sec.~\ref{sec:train_perform}) we pick a window of $V$ columns to form the mini-batch on each stochastic gradient descent~(SGD) optimization step. To ensure a sufficiently large mini-batch of independent examples, when forming the data-sets, we split one full DAC load into two parts, treated independently by the learning algorithm, e.g. $\mathbf{y}_{1,1\ldots T} \mathbf{y}_{2,1\ldots T} \mathbf{y}_{3,1\ldots T} \mathbf{y}_{4,1\ldots T}$ and $\mathbf{y}_{5,1\ldots T}\mathbf{y}_{6,1\ldots T} \mathbf{y}_{7,1\ldots T} \mathbf{y}_{8,1\ldots T}$. Note that this results in only negligible performance degradation since $4T\gg V$. We verified that convergence of the loss is achieved within one epoch of the training data, which we then used as a stopping criterion. The remaining sequences with $i\in\{721,\ldots,Z\}$, indepent from the training set are used in the testing phase. 

\subsection{Data Collection for the PAM Systems}\label{sec:PAM_data}
We consider the conventional two- and four-level PAM and investigate deep learning approaches for detection based on recurrent and feed-forward neural networks as well as nonlinear Volterra equalization. At the transmitter we use a Mersenne twister to generate PAM2 or PAM4 symbols which we accordingly scaled to $\mathcal{M}=\{0;\pi/4\}$ or $\mathcal{M}=\{0;\pi/12;\pi/6;\pi/4\}$, respectively, enabling operation in the linear region of the MZM. Note that Gray-coded bit-to-symbol mapping was explicitly assumed in the case of PAM4. The sequence of symbols is pulse-shaped at $n\!=\!2$ Sa/sym by a 0.25 roll-off raised cosine filter. It is then applied to the experimental test-bed. We performed $Z\!=\!100$ loads of the DAC, each of them equivalent to a sequence of $T=249 750$ PAM symbols $m_{i,j}$, $i\in\{1,\ldots,Z\}$ and $j\in\{1,\ldots,T\}$. The sequences of received symbols $\mathbf{y}_{i,j}$ after propagation through the test-bed link were utilised together with the labels $m_{i,j}$ for optimization of the DSP parameters and testing. Sequences $i\in\{1,\ldots,90\}$ were used to form the training sets
\begin{equation*}
\mathbf{D}^{\textnormal{PAM}}:= 
\begin{bmatrix}
\mathbf{y}_{1,1} & \ldots & \mathbf{y}_{1,T/2}  \\
\mathbf{y}_{1,T/2+1} & \ldots & \mathbf{y}_{1,T}  \\
\vdots & \ddots & \vdots \\
\mathbf{y}_{90,1} & \ldots & \mathbf{y}_{90,T/2}  \\
\mathbf{y}_{90,T/2+1} & \ldots & \mathbf{y}_{90,T}  \\
\end{bmatrix}
\end{equation*}
\begin{equation*}
\mathbf{L}^{\textnormal{PAM}}:= 
\begin{bmatrix}
m_{1,1} & \ldots & m_{1,T/2}  \\
m_{1,T/2+1} & \ldots & m_{1,T}  \\
\vdots & \ddots & \vdots \\
m_{90,1} & \ldots & m_{90,T/2}  \\
m_{90,T/2+1} & \ldots & m_{90,T}  \\
\end{bmatrix}
\end{equation*}
where $\mathbf{D}^{\textnormal{PAM}}\!\in\!{\mathbb{R}}^{180\times T\cdot n/2}$ consists of the received elements $\mathbf{y}_{i,j}$ and $\mathbf{L}^{\textnormal{PAM}}\!\in\!{\mathbb{R}}^{180\times T/2}$ has the elements $m_{i,j}$. During the training of the PAM systems we pick column-wise with a window of $W$ from the data to form a mini-batch. Similar to the auto-encoder training one full DAC load is split into two independently treated parts, e.g. $\mathbf{y}_{1,1} \ldots \mathbf{y}_{1,T/2}$ and
$\mathbf{y}_{1,T/2+1} \ldots \mathbf{y}_{1,T}$, ensuring an identical mini-batch size. Again, one epoch over the data was performed. Testing is performed on the remaining sequences $i\in\{91,\ldots,100\}$.

\section{Digital Signal Processing Schemes and Performance}\label{sec:DSP_schemes}
\begin{figure}[b!]\vspace*{-3ex}
\centering
\includegraphics[width=\columnwidth, keepaspectratio=true]{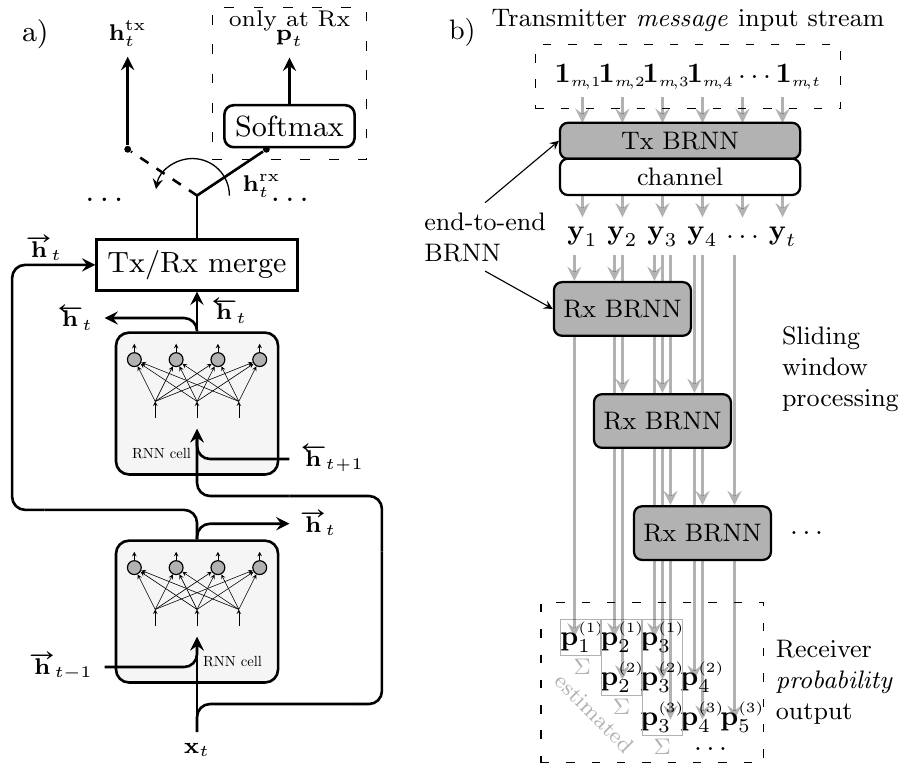}\vspace*{-2ex}
\caption{\label{fig:bidirectional_sliding_schematics} Schematic of a) bidirectional RNN processing and b) sliding window sequence estimation algorithm.}
\end{figure}
The IM/DD transmission is distorted by chromatic dispersion, which introduces ISI from both preceding and succeeding symbols, and square-law detection by a photo-diode~\cite{Agrawal}. The joint effects of these phenomena render the communication channel nonlinear with memory, necessitating processing to recover data sequences. However, due to the lack of optimal, computationally feasible algorithms, carefully chosen approximations are required. In this section we investigate classical DSP as well as deep learning-based solutions.

\subsection{Sliding Window BRNN Auto-encoder}\label{sec:SBRNN_transceiver}
An efficient auto-encoder implements the complete chain of transmitter, channel and receiver as an end-to-end deep ANN, whose parameters are optimized in a single, joint learning process~\cite{O'Shea}. In particular, for optical IM/DD, a bidirectional recurrent neural network (BRNN) can be used to handle the memory of the nonlinear channel~\cite{Karanov_2,Karanov_3}. As proposed in~\cite{Farsad}, the BRNN is combined with a powerful receiver algorithm for sequence estimation.
\subsubsection{BRNN processing}
The schematic of the BRNN building block, used at both transmitter and receiver, is visualized in Fig.~\ref{fig:bidirectional_sliding_schematics}-a). In the forward direction, the input $\textnormal{\textbf{x}}_{t}$ at time $t$ is processed by the recurrent cell together with the previous output $\overrightarrow{\textnormal{\textbf{h}}}_{t-1}$ to produce the updated output $\overrightarrow{\mathbf{h}}_{t}=\alpha_{\textnormal{Tx/Rx}}\left(\mathbf{W}\begin{pmatrix} \mathbf{x}_t^T & \overrightarrow{\mathbf{h}}_{t-1}^T\end{pmatrix}^T +\mathbf{b}\right)$, where ${}^{T}$ denotes the matrix transpose, $\mathbf{W}_{}\in{\mathbb{R}}^{n \times (M+n)}$ and $\mathbf{b}_{}\in{\mathbb{R}}^{n}$ (transmitter) or $\mathbf{W}_{}\in{\mathbb{R}}^{2M \times (n+2M)}$ and $\mathbf{b}_{}\in{\mathbb{R}}^{2M}$ (receiver) are the weight matrix and bias vector, respectively, and $\alpha_{\textnormal{Tx/Rx}}$ is the activation function. In the backward direction $\textnormal{\textbf{x}}_{t}$ is instead combined with $\overleftarrow{\textnormal{\textbf{h}}}_{t+1}$ to yield $\overleftarrow{\textnormal{\textbf{h}}}_{t}$. The output of the transmitter BRNN at time $t$ is
$\mathbf{h}_{t}^{\textnormal{tx}} = \frac{1}{2}\left({\overrightarrow{\mathbf{h}}}_{t}+{\overleftarrow{\mathbf{h}}}_{t}\right)$, while at the receiver $\mathbf{h}_{t}^{\textnormal{rx}} = \begin{pmatrix} \overrightarrow{\mathbf{h}}_t^T & \overleftarrow{\mathbf{h}}_{t}^T\end{pmatrix}^T$. To enable approximately linear MZM operation, the transmitter activation is the clipping function $\alpha_{\text{Tx}}(\mathbf{a})=\alpha_{\tiny{\text{ReLU}}}\left(\mathbf{a}\right)-\alpha_{\tiny{\text{ReLU}}}\left(\mathbf{a}-\frac{\pi}{4}\right)$~\cite{Karanov_1}. At the receiver, we employ the ReLU activation $\alpha_{\text{Rx}}(\mathbf{a})=\alpha_{\tiny{\text{ReLU}}}\left(\mathbf{a}\right)$~\cite{Goodfellow}. After the ReLU nodes, a \emph{softmax} layer is applied to obtain probability vectors at the receiver according to \mbox{$\mathbf{p}_t = \mathop{\textrm{softmax}}(\mathbf{W}_{\text{softmax}}\mathbf{h}_{t}^{\textnormal{rx}} + \mathbf{b}_{\text{softmax}})$} with $\mathbf{p}_t\in{\mathbb{R}}^{M}$, $\mathbf{W}_{\text{softmax}}\in{\mathbb{R}}^{M\times 4M}$ and $\mathbf{b}_{\text{softmax}}\in{\mathbb{R}}^{M}$.

In the  auto-encoder setting, the BRNN transmitter takes as an input a sequence of independently drawn messages $m$, represented as a stream of one-hot vectors $(\ldots,\mathbf{1}_{m,t-1},\mathbf{1}_{m,t},\mathbf{1}_{m,t+1},\ldots)$, with $\mathbf{1}_{m,t}\in{\mathbb{R}}^{M}$ (with elements ``1'' at position $m$ and ``0'' elsewhere). It encodes them into a sequence of transmit blocks of $n$ samples $(\ldots,\mathbf{h}_{t-1}^{\textnormal{tx}},\mathbf{h}_t^{\textnormal{tx}},\mathbf{h}_{t+1}^{\textnormal{tx}},\ldots)$. This sequence is sent through the communication channel. The receiver processes the noisy samples $(\ldots,\mathbf{y}_{t-1},\mathbf{y}_t,\mathbf{y}_{t+1},\ldots)$ to obtain $(\ldots,\mathbf{h}_{t-1}^{\textnormal{rx}},\mathbf{h}_t^{\textnormal{rx}},\mathbf{h}_{t+1}^{\textnormal{rx}},\ldots)$ and then probability vectors $(\ldots,\mathbf{p}_{t-1},\mathbf{p}_t,\mathbf{p}_{t+1},\ldots)$ via softmax.

\subsubsection{Sliding window sequence estimation algorithm} \label{sec:sliding_descr}
In principle, one could apply the BRNN on the full sequence. However, to improve receiver latency, we employ a windowed processing, as proposed in~\cite{Farsad}. Figure~\ref{fig:bidirectional_sliding_schematics}-b) shows a schematic of the end-to-end processing and estimation of data sequences. For a given sequence of $T$ messages, the transmitter BRNN encodes the full stream of input one-hot vectors $\mathbf{1}_{m,1}, \ldots, \mathbf{1}_{m,T}$. The obtained waveform is then subject to the channel, yielding the sequence of received blocks of samples $\mathbf{y}_1, \ldots, \mathbf{y}_{T}$. At a time $t$, the receiver BRNN processes the window of $W$ received blocks $\mathbf{y}_t, \ldots, \mathbf{y}_{t+W-1}$, transforming them into $W$ probability vectors $\mathbf{p}_{t}^{(t)},\ldots, \mathbf{p}_{t+W-1}^{(t)}$. Then it slides one time slot ahead to process the blocks  $\mathbf{y}_{t+1}, \ldots, \mathbf{y}_{t+W}$. The final output probability vectors  are given by
\begin{equation}
\label{eq:Sliding_window_2}
\mathbf{p}_{i}=\sum\limits_{k=0}^{\max(W,i)-1}a^{(k)}_i\mathbf{p}_{i}^{(i-k)}
\end{equation}
where we set $a^{(q)}_i = [\max(W,i)]^{-1}$. We refer the interested reader to~\cite[III-B]{Karanov_4} for a study on optimizing the weight coefficients $a^{(q)}_i$, yielding additional small gains compared to the uniform assumption followed in this work. Decision on the transmitted message can be made using the probability output of the estimation algorithm, i.e. $\tilde{m}_{i}=\argmax(\mathbf{p}_{i})$. We count a block (symbol) error when $m_{i}\neq \tilde{m}_{i}$. The block error rate (BLER) for the transmitted sequence is thus given by
\begin{equation}
\text{BLER}=\frac{1}{T}\sum\limits_{i=1}^{T}{\mathbbm{1}}_{\left\{m_{i}\neq \tilde{m}_{i}\right\}},
\end{equation}
where $\mathbbm{1}_{\{\cdot\}}$ denotes the indicator function, equal to 1 when the argument is satisfied and 0 otherwise. Using the method described in~\cite[Sec. III-C]{Karanov_3}, we perform optimization of the bit-to-symbol
mapping to obtain the BER. The reported BER is the average BER over all test sequences.

\begin{figure}[t!]
\centering
\includegraphics[width=\columnwidth, keepaspectratio=true]{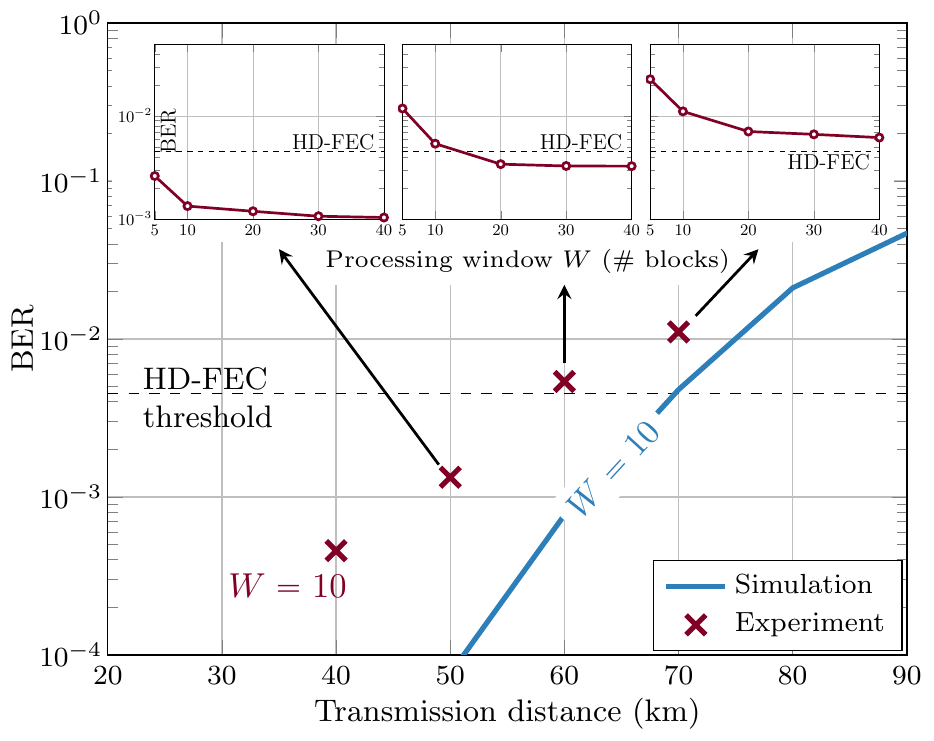}\vspace*{-1.5ex}
\caption{\label{fig:Window_results} BER versus distance for the SBRNN auto-encoder trained on a channel model. Inset figures show the BER vs. window size $W$ at 50, 60 and 70\,km.}\vspace*{-2ex}
\end{figure}
\begin{figure}[t!]
\centering
\includegraphics[width=0.9\columnwidth, keepaspectratio=true]{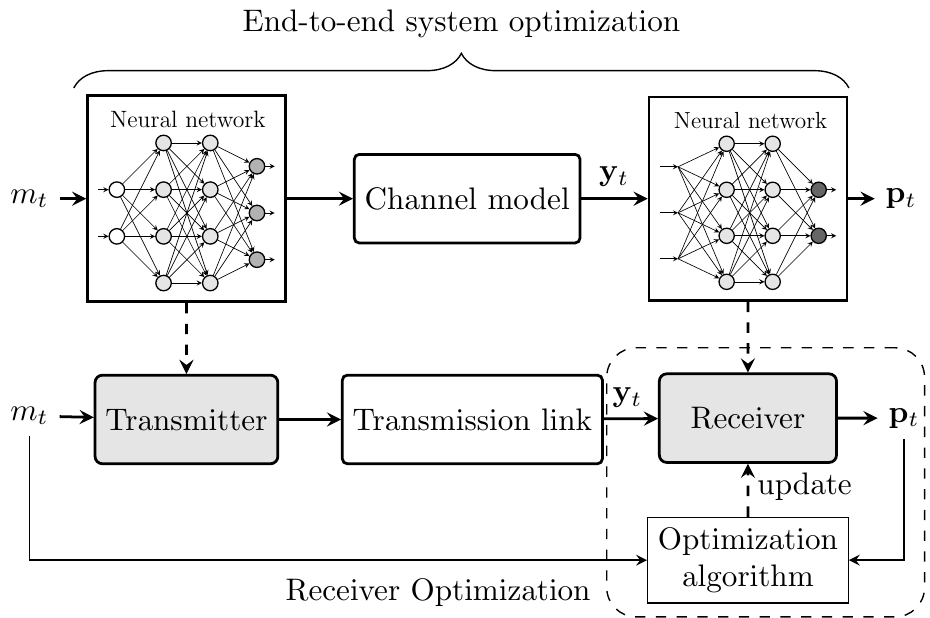}
\caption{\label{fig:receiver_opt_schematic} Receiver optimization using the collected data from measurements.}\vspace*{-3ex}
\end{figure}
\subsubsection{Training and performance}\label{sec:train_perform}
First, the end-to-end system learning is performed following~\cite[Sec.~III-A]{Karanov_3}) on a numerical link model identical to~\cite[Sec.~2.1]{Karanov_2}. The optimized transceiver is applied ``as is'' to the experimental test-bed. Figure~\ref{fig:Window_results} shows the performance of this system for a fixed sliding window size of $W\!=\!10$. After 50\,km, a BER of $1.3\cdot10^{-3}$ is achieved, while at 60\,km it is slightly above the $4.5\cdot 10^{-3}$ HD-FEC threshold~\cite{Wang} (6.7\% overhead). A simple method for performance improvement at longer distances is to increase the window in the sequence estimation algorithm, thus compensating for more of the ISI. The inset figures examine the BER at 50, 60 and 70\,km as a function of $W$. We see that at 60\,km, the system can still operate below the HD-FEC threshold with $W\!=\!20$. In an excellent agreement with the simulation~\cite{Karanov_2}, the experiment shows diminishing gains from further increase of $W$ since noise and non-linearities become more dominant when the ISI is compensated.

A different approach for improving the system performance is shown in~Fig.~\ref{fig:receiver_opt_schematic}. It uses the collected experimental data for receiver optimization, allowing to reduce the penalty stemming from the discrepancy between simulation model and the actual transmission link~\cite{Doerner}. Training elements are the pairings of an input message $m_t$ from $\mathbf{L}^{\textnormal{AE}}$ and the corresponding received block of samples $\mathbf{y}_{t}$ from $\mathbf{D}^{\textnormal{AE}}$ (see Sec.~\ref{sec:autoenc_data}). The receiver parameters $\boldsymbol{\theta_{\textnormal{Rx}}}$ are adjusted via SGD,
minimizing the average loss $L$ over a mini-batch $S$ of the training data, given by
\begin{equation}
L(\boldsymbol{\theta_{\textnormal{Rx}}})=\frac{1}{|S|}\sum\limits_{(m_t,\mathbf{y}_{t})\in{S}}\ell(m_t, f_{\textrm{rec},t}(\ldots,\mathbf{y}_{t},\ldots)),
\end{equation}
where $\ell(\mathbf{x},\mathbf{y})\!=\!-\sum_{i}x_i\log(y_i)$ is the cross entropy and $f_{\textrm{rec}}$ is the receiver ANN function.
More specifically, training is performed in the following way: At the beginning, the outputs ${\overrightarrow{\mathbf{h}}}_{t-1}$ and ${\overleftarrow{\mathbf{h}}}_{t+1}$ in the forward and backward passes of the BRNN are initialized to $\mathbf{0}$. At an optimization step $s$, the mini-batch of elements\footnote{We use the MATLAB notation $\mathbf{A}[:,i\!:\!j]$ to denote extracting the $j-i+1$ columns of column indices $i, i+1, \ldots, j$ from $\mathbf{A}$.} $\mathbf{D}^{\textnormal{AE}}[:,s\!:\!(s+V)]$ is processed by the receiver to obtain corresponding output probability vectors, where $V$ is the training window which we fixed to 10. The loss between the input messages $\mathbf{L}^{\textnormal{AE}}[:,s\!:\!(s+V)]$ and probability outputs is computed and averaged before completing a single iteration of the optimization algorithm. After optimization, the system performance is evaluated using the testing set and reported in Sec.~\ref{sec:Performance}. Note that transmitter optimization from experimental data has been implemented for optical fiber systems in~\cite{Karanov_5}. To avoid the additional complexity, we do not consider such an optimization in this paper.

\subsection{Sliding Window BRNN PAM Receiver}\label{sec:PAM_SBRNN}
This system combines the PAM transmitter with the SBRNN receiver. The receiver is implemented following an identical procedure to Sec.~\ref{sec:SBRNN_transceiver}. Differently, here we used the elements of $\mathbf{D}^{\textnormal{PAM}}$ and $\mathbf{L}^{\textnormal{PAM}}$ for training. At an iteration $s$, the mini-batch of elements $\mathbf{D}^{\textnormal{PAM}}[:,s\!:\!(s+V)]$ is processed by the BRNN to obtain the corresponding output probability vectors. We compute the average cross entropy loss between the input messages $\mathbf{L}^{\textnormal{PAM}}[:,s:(s+V)]$ and these outputs before completing an optimization step of the ANN parameters. We fixed $V\!=\!61$, such that the scheme is trained with a processing memory identical to the auto-encoder.

\subsection{Sliding Window FFNN PAM Receiver}\label{sec:PAM_SFFNN}
\begin{figure}[t!]
\centering
\includegraphics[width=0.75\columnwidth, keepaspectratio=true]{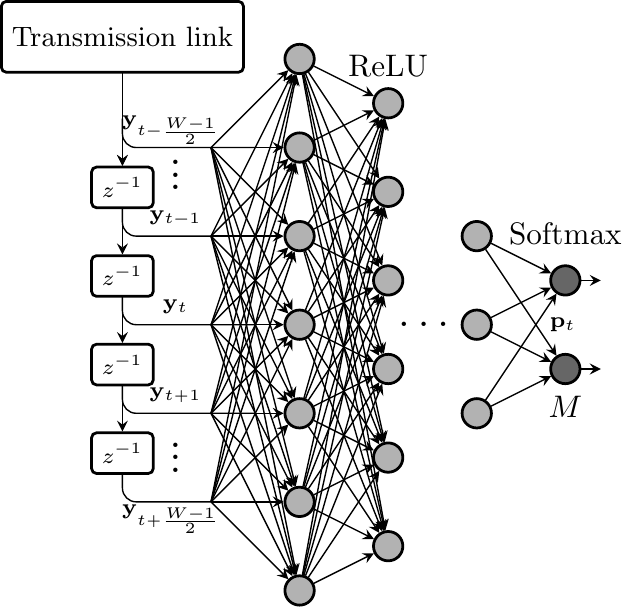}\vspace*{-1ex}
\caption{\label{fig:SFFNN_schematic} Schematic of the sliding window FFNN receiver for PAM symbols.}\vspace*{-2ex}
\end{figure}
\begin{figure}[t!]
\centering
\includegraphics[width=0.8\columnwidth, keepaspectratio=true]{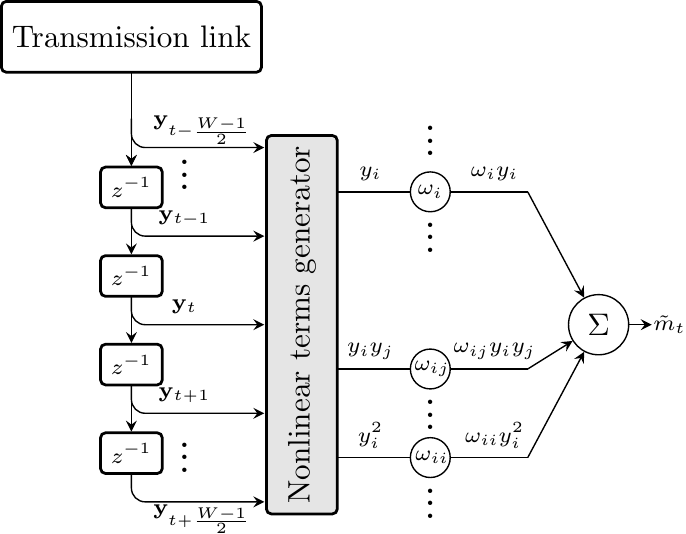}\vspace*{-1ex}
\caption{\label{fig:Volterra_schematic} Schematic of the nonlinear Volterra equalizer for PAM symbols.}\vspace*{-2ex}
\end{figure}
In this scheme, a PAM sequence is transmitted through the link and the received samples are fed to a multi-layer feed-forward ANN, as shown in Fig.~\ref{fig:SFFNN_schematic} estimating the center symbol of the sequence. More specifically, the transmitted message $m_t$ at time $t$ is detected by processing the train of $W$ received symbols of $n\!=\!2$ samples $(\mathbf{y}_{t-\frac{W-1}{2}},\ldots,\mathbf{y}_{t},\ldots,\mathbf{y}_{t+\frac{W-1}{2}})$. In this receiver, the processing memory dictates the size of the layers, the first one having parameters $\mathbf{W}\!\in\!{\mathbb{R}}^{W \cdot n\times 4W \cdot n}$ and $\mathbf{b}\!\in\!{\mathbb{R}}^{4W \cdot n}$. Subsequently, 6 hidden layers are employed before $m_t$ is estimated. The number of nodes on each of these is given by $\left\lfloor 4W\cdot n/(2^{i-1})\right\rfloor$, where $i$ is the hidden layer index. The ReLU activation is applied on all except the final layer, where \emph{softmax} computes a probability vector for the transmitted PAM2/4 symbol with $\mathbf{p}_t\!\in\!{\mathbb{R}}^{2/4}$. Afterwards, we slide the processing window by one position ahead to estimate the next symbol in the sequence. Note that a similar approach has been considered previously in~\cite{Houtsma,Karanov_2}. Training is performed in the following way: at an iteration $s$, the mini-batch of elements $\mathbf{D}^{\textnormal{PAM}}[:,s\!:\!(s+W)]$ is processed by the FFNN to obtain the corresponding output probability vectors. We compute the average cross entropy between these outputs and the input messages $\mathbf{L}^{\textnormal{PAM}}[:,s+\frac{W-1}{2}]$ and complete an optimization step of the parameters via SGD.
\subsection{Nonlinear Volterra Equalizer for PAM Transmission}\label{sec:PAM_Volterra}
Volterra equalizers, shown schematically in Fig.~\ref{fig:Volterra_schematic}, have been widely considered for IM/DD links based on PAM~\cite{Stojanovic}. To compensate for linear and nonlinear ISI, we use a second order Volterra filter, whose  output can be expressed as
\begin{equation}
    \begin{split}
        \tilde{m}_t & =\omega_{dc}+\sum\limits_{q_1=-\frac{W-1}{2}\cdot n}^{\frac{W-1}{2}\cdot n}\omega_{q_1}\cdot y_{t+q_1}\\ & +\sum\limits_{q_2=-\frac{W_1-1}{2}\cdot n}^{\frac{W_1-1}{2}\cdot n}\sum\limits_{q_3=k_2}^{\frac{W_1-1}{2}\cdot n}\omega_{q_2,q_3}\cdot y_{t+q_2}\cdot y_{t+q_3},
    \end{split}
\end{equation}
where $W$ and $W_1$ denote the symbol memory in the first and second order series, respectively, and $\omega_{i}$ and $\omega_{i,j}$ are the first and second order coefficients. Similar to the other PAM schemes, the algorithm is processing the neighborhood of $W$ received symbols of $n\!=\!2$ samples $(\mathbf{y}_{t-\frac{W-1}{2}},\ldots,\mathbf{y}_{t},\ldots,\mathbf{y}_{t+\frac{W-1}{2}})$, thus including the ISI from pre- and post-cursor samples in the detection of the current input. To capture identical amount of memory, we fix $W\!=\!61$, while we choose $W_1\!=\!21$ to keep the complexity low. The filter coefficients are optimized using the MATLAB \verb|fitlm|
function for linear and polynomial regression. Training is performed on a randomly chosen pair of transmitted sequence $\mathbf{L}^{\textnormal{PAM}}[i,:]$ and received samples \mbox{$\mathbf{D}^{\textnormal{PAM}}[i,:]$}.

\begin{figure}[t!]
\centering
\includegraphics[width=\columnwidth, keepaspectratio=true]{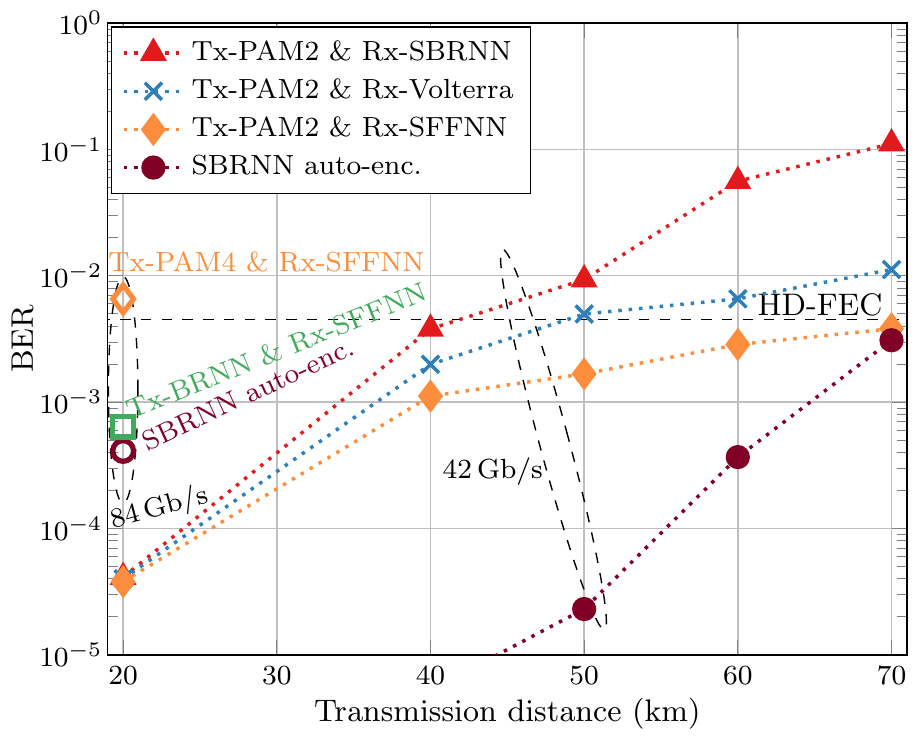}\vspace*{-1.5ex}
\caption{\label{fig:Performance_results} BER as a function of distance for 42\,Gb/s and 84\,Gb/s systems employing DSP schemes trained on experimental data.}\vspace*{-2ex}
\end{figure}
\subsection{Performance of the DSP optimized on experimental data}\label{sec:Performance}
To compare the performance of the DSP, we fix the number of bits simultaneously processed by each receiver algorithm. Thus, for the auto-encoder we set $W\!=\!10$~(60 bits), while for the PAM2 systems $W\!=\!61$~(61 bits). To allow for a better compensation in the case of PAM4, we kept $W=61$~(122 bits). Figure~\ref{fig:Performance_results} shows the BER of all experimental systems at different distances. We see that for shorter lengths and 42\,Gb/s, the SBRNN auto-encoder has a BER much lower than the PAM schemes. In terms of system reach, it allows transmission up to 70\,km below HD-FEC, similar to the PAM2\&SFFNN. Both schemes outperformed the Volterra equalizer. The results show the SFFNN as a viable receiver solution for the conventional PAM. For further investigation, we employ it with PAM4 in a 84\,Gb/s system. Moreover, we combined the SFFNN receiver in an auto-encoder setting with the BRNN transmitter. The results show that the PAM4\&SFFNN cannot transmit below the HD-FEC at 20\,km, while the Tx-BRNN\&Rx-SFFNN auto-encoder is significantly superior, achieving a BER close to the SBRNN. Nevertheless, it is worth mentioning that, unlike the SBRNN, the number of parameters in the SFFNN receiver increases rapidly with the processing memory.
\section{Conclusions}\label{sec:Conclusions}
We investigated the experimental performance of deep learning in end-to-end as well as receiver-only application for DSP in short reach optical fiber links. Our results show that 42\,Gb/s can be transmitted at up to 60\,km below a common HD-FEC threshold by a simple SBRNN auto-encoder which learns transceiver parameters based on a model. The reach can be enhanced to 70\,km by optimization of the receiver on experimental data. Interestingly, 42\,Gb/s systems based on PAM2 and receiver-only DSP by an SFFNN covered similar distances, outperforming nonlinear Volterra equalization. In addition to the SBRNN, we examined an auto-encoder based on BRNN transmitter and SFFNN receiver. Both configurations allowed to transmit 84\,Gb/s at 20\,km.  Our experiment highlights deep learning as a viable DSP solution for low-cost optical communications to increase reach or enhance the data rate at shorter distances.

\end{document}